\documentclass{Interspeech2024}

\usepackage{graphicx}
\usepackage{amsmath}
\usepackage{amssymb}
\usepackage{amsthm}
\usepackage{bm,bbold}
\usepackage{nicefrac}
\usepackage{multicol}
\usepackage{pifont}
\usepackage[symbol]{footmisc}
\newtheorem{thm}{Theorem}[section]

\newtheorem*{prop*}{Proposition}

\newtheorem*{cor*}{Corollary}

\theoremstyle{remark}

\usepackage{mathtools}
\usepackage{siunitx}
\usepackage{enumitem}
\usepackage{adjustbox}
\usepackage{stackengine}
\usepackage{url}
\usepackage{microtype}
\usepackage[dvipsnames]{xcolor}
\usepackage{hyperref}
\usepackage{booktabs}
\usepackage{colortbl}
\newcommand\blfootnote[1]{%
  \begingroup
  \renewcommand\thefootnote{}\footnote{#1}%
  \addtocounter{footnote}{-1}%
  \endgroup
}

\definecolor{lightgray}{gray}{0.95}

\newcommand{\myPhi}{\Phi}

\newcommand{\myPsi}{\Psi}
\newcommand{\mypsi}{\psi}

\newcommand{\myw}{w}

\newcommand{\myx}{x}

\usepackage[acronym]{glossaries}
\newacronym{SNR}{SNR}{signal-to-noise ratio}

\newcommand{\RR}{\mathbb{R}}

\definecolor{darkviolet}{rgb}{0.9,0,0.9} 

\interspeechcameraready

\title{Hold Me Tight: Stable Encoder--Decoder Design for Speech Enhancement}

\name[affiliation={1}]{Daniel}{Haider$^*$}
\name[affiliation={1}]{Felix}{Perfler$^*$}
\name[affiliation={2}]{Vincent}{Lostanlen}
\name[affiliation={3}]{Martin}{Ehler}
\name[affiliation={1}]{Peter}{Balazs}

\address{
  $^1$Acoustics Research Institute, Austrian Academy of Sciences, Austria\\
  $^2$Nantes Université, École Centrale Nantes, CNRS, LS2N, France \\
  $^3$University of Vienna, Faculty of Mathematics, Austria}
\email{daniel.haider@oeaw.ac.at, felix.perfler@oeaw.ac.at}

\keywords{Hybrid filterbanks, stabilization, tight frames, encoder, reconstruction, speech enhancement}

\begin{document}

\maketitle

\begin{abstract}
Convolutional layers with 1-D filters are often used as frontend to encode audio signals. Unlike fixed time--frequency representations, they can adapt to the local characteristics of input data.
However, 1-D filters on raw audio are hard to train and often suffer from instabilities.
In this paper, we address these problems with hybrid solutions, i.e., combining theory-driven and data-driven approaches. 
First, we preprocess the audio signals via a auditory filterbank, guaranteeing good frequency localization for the learned encoder.
Second, we use results from frame theory to define an unsupervised learning objective that encourages energy conservation and perfect reconstruction. Third, we adapt mixed compressed spectral norms as learning objectives to the encoder coefficients. 
Using these solutions in a low-complexity encoder--mask--decoder model significantly improves the perceptual evaluation of speech quality (PESQ) in speech enhancement.
\end{abstract}

\section{Introduction}
Time--frequency transforms, such as the short--time Fourier transform (STFT) and the constant-Q transform (CQT), have long served for analysis and synthesis of audio \cite{Pejman2017}.
More recently, neural networks have started to outperform these classical methods \cite{Vieting2021, sainath2015learning}. 
While both approaches are being used in applications, they come with different pros and cons.
On the one hand, fixed transforms are controllable and interpretable; but the chosen 
time--frequency resolution may be suboptimal for the task at hand.
On the other hand, learnable transforms have the potential to adapt to the short--term properties of the data; but they remain difficult to train, less interpretable, and potentially unstable.

The diverging opinions on the relative merits of the two approaches are particularly evident in encoder--mask--decoder models \cite{lou2019convtasnet}.
As of today, the mask is typically estimated by a neural network that is trained on the coefficients of the encoded input signals \cite{braun2020data,Li2021RealtimeMS,ditter2020gammatasnet}.
For the encoder--decoder design, there are two main paradigms: time--frequency domain methods use fixed time--frequency transforms (e.g., STFT \cite{braun2020data}, mel spectrogram, Gammatone filters \cite{ditter2020gammatasnet}), and time-domain methods process the signal waveforms directly via a convolutional layer with 1-D filters that is optimized together with the model parameters, also known as filterbank learning \cite{sainath2015learning,lou2019convtasnet}.
The ongoing competition between these paradigms \cite{heitkaemper20tasnet} indicates that the optimal way of encoding audio signals remains an unsolved problem \cite{dorfler2020basic}.

\begin{figure}[t]
    \centering
    \includegraphics[width =\linewidth]{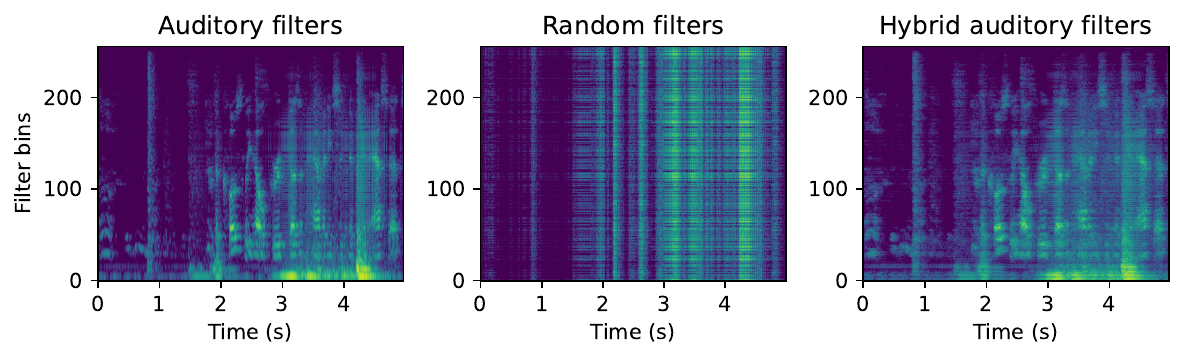}
    \caption{The log magnitude responses of three encoders for the same speech signal. Left to right: Auditory filterbank, random filterbank, and hybrid filterbank as channel--wise composition of the previous two. While the random responses are hard to interpret, the hybrid responses are comparable to the fixed ones with the possibility to be fine--tuned in a data--driven manner.}
    \label{fig:resp}
\end{figure}

A family of models, so-called \emph{hybrid}, aim to combine feature engineering with feature learning.
These models rely on domain knowledge so as to reduce optimization to certain properties of filters, such as center frequencies, bandwidth, and gain \cite{ravanelli2018sincnet, zeghidour2021leaf, seki2017fblearning}.
Recently, a hybrid architecture known as multiresolution neural network (MuReNN) was proposed to fit auditory filterbanks from data \cite{lostanlen2023murenn}.
In MuReNN, small filterbanks are learned on different resolution levels by using a wavelet decomposition of the input signal. This can be implemented as level-wise convolutions of wavelet filters with trainable filters.

In our work, we propose a comparable hybrid construction specifically for speech processing.
While MuReNN relies on a discrete wavelet transform, we use an auditory filterbank.
This choice prioritizes perceptually significant aspects of speech, such as the need for high temporal resolution at lower frequencies. The hybrid construction then allows for further refinement of the corresponding signal representations.
When combined with a mixed compressed spectral learning objective that respects these representations, this configuration provides a flexible and powerful setup for hybrid filterbank learning.

To facilitate the reconstruction of the enhanced signal from the encoder domain, a suitable decoder is essential. If the encoder--decoder pair yields perfect reconstruction (without a mask) the decoder is called dual to the encoder. If the encoder is dual to itself, it is said to be \textit{tight}. This case is particularly advantageous as it simplifies reconstruction and ensures stability for the encoder through norm preservation \cite{Boelcskei01noisereduction,yu2008denoise,badokowto13}, which improves the robustness against noise and adversarial examples \cite{hasannasab2020parseval,cisse2017parseval}. By incorporating a measure of tightness of the encoder into the learning objective, we can use the hybrid filterbank in a encoder--decoder setting without the need for computing a dual.

In summary, this paper presents a complete framework for training hybrid filterbanks in an encoder--decoder setting for speech--related tasks by (i) conceptually fusing auditory filterbank design with classical filterbank learning, (ii) stabilizing encoders during training by promoting tightness, and (iii) adapting the learning objective to the encoder coefficient domain.\blfootnote{$^*$ Equal contribution. Code available at\\ \url{https://github.com/felixperfler/Stable-Hybrid-Auditory-Filterbanks}.}


\begin{figure*}[t]
    \centering
    \includegraphics[width =\linewidth]{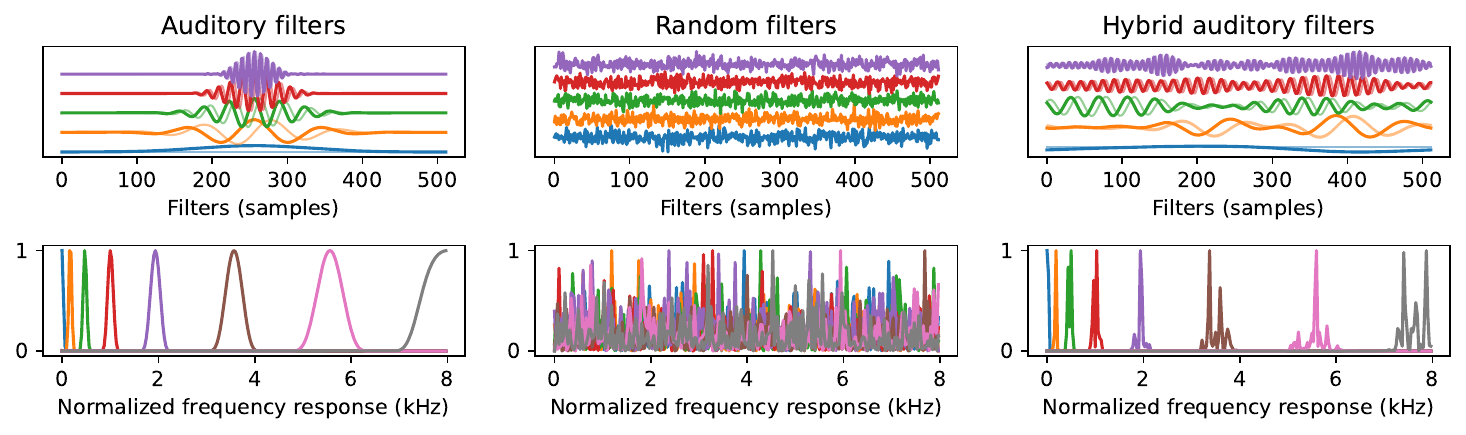}
    \caption{Selections of real and imaginary parts of filters (top) and their frequency responses (bottom) from three different filterbanks. From left to right: An auditory filterbank with center frequencies uniformly on the mel scale, a random filterbank with $\sigma^2 = (TJ)^{-1}$, and a hybrid auditory filterbank as the channel-wise composition of the previous two. Different filters are displayed with different colors.
    }
    \label{fig:filters}
\end{figure*}

\section{Learning Tight Hybrid Filterbanks with Inductive Auditory Bias}
This section establishes the mathematical foundation for our methodology.
Let $\myx \in \RR^N$ be an audio signal. A convolutional layer $\myPhi$ with 1-D kernels $\myw_j\in \RR^T$, $T\leq N$ decomposes $\myx$ into $J>1$ subbands via convolution, represented as the array
\begin{equation*}
    \left(\myPhi \myx\right)[n,j] = (\myx \ast \myw_j)[n] = \sum_{k=0}^{T-1}\myw_j[k]\myx[(n-k) \operatorname{mod} N],
\end{equation*}
also referred to as \textit{responses} of $\myPhi$ for $x$.
In the context of classical signal processing this corresponds to an oversampled finite impulse response (FIR) filterbank \cite{boelcskei1998filterbanks}.
Besides all common linear time--frequency transforms, such as the STFT and the CQT, adaptive or adapted auditory--related time--frequency representations can be envisioned and implemented in this way as well \cite{Necciari:2013a,necciari2018audlets}. One obstacle to a successful and functional implementation of such customized filterbanks is stability. 

\subsection{Tight Filterbank Frames}
A filterbank $\myPhi$ forms a \textit{frame} for $\mathbb{R}^N$ if there are positive constants $A\leq B$ such that
\begin{equation}\label{eq:frame}
    A\cdot \Vert \myx \Vert^2 \leq
\Vert \myPhi \myx \Vert^2 \leq
B\cdot \Vert \myx \Vert^2
\end{equation}
holds for any $\myx \in \RR^N$ \cite{christensen2002intro}. The numbers $A,B$ are called the frame bounds. This Lipschitz-type inequality guarantees that the filterbank decomposition is invertible and well-conditioned, i.e., \textit{stable}. The optimal bounds (largest $A$, smallest $B$) in \eqref{eq:frame} are given by the smallest and largest eigenvalues of the associated \emph{frame operator} $\myPhi^\top \myPhi$, where $\myPhi^\top$ denotes the transposed filterbank of $\myPhi$.
These values determine the numerical stability of $\myPhi$ via the condition number $\kappa = B/A$ \cite{casazza2012finite}. Hence, a filterbank with $A=B$ has optimal stability properties and is called \textit{tight}.
For a tight filterbank $\myPhi$, the following are equivalent \cite{casazza2012finite}.
\begin{enumerate}
\setlength{\itemindent}{2.5em}
    \item[(i)] $\Vert \myPhi \myx \Vert^2 = A\cdot\Vert \myx \Vert^2$ for all $\myx\in\RR^N$,
    \item[(ii)] $\myPhi^\top \myPhi = A\cdot \mathbb{1}_N$
    \item[(iii)] $\kappa= B/A=1$.
\end{enumerate}
Property (i) states that the filterbank is norm-preserving. This is advantageous as the energy level of the encoder responses is always under control and different signal parts contribute equally. In particular, this makes $\myPhi$ robust to small perturbations, which is a crucial property in the context of adversarial examples \cite{cisse2017parseval}.

Property (ii) is especially interesting in an encoder-decoder regime: If the encoder filterbank $\Phi$ is tight, then the transposed filterbank $\myPhi^\top$ as a decoder yields perfect reconstruction \cite{balazs2017framespsycho}. Hence, no inverse decoder has to be computed or learned, which decreases the computational complexity significantly.

Property (iii) coincides with the classical definition of optimal stability of the linear operator associated with $\myPhi$ from numerical linear algebra. To make encoder filterbanks with trainable weights benefit from (i) and (ii), we propose to minimize $\kappa$ during training in parallel with the objective function (Sec. \ref{sec:stab}).

\subsection{Encoder Design: Hybrid Auditory Filterbanks}
Following the idea of multiresolution neural networks \cite{lostanlen2023murenn} (MuReNN), we compose the filters from a fixed filterbank with trainable filters via convolution. Letting $\myPsi$ denote the fixed filterbank with filters $\mypsi_i$ and $\myPhi$ the filterbank with trainable filters $\myw_j$, then we define the trainable hybrid filterbank $\myPhi_{\myPsi}$ as the filterbank with filters $(\myw_j \ast \mypsi_j)$ for every $j$. Hence, any signal $\myx$ is decomposed as
\begin{equation}\label{eq:aud_bias}
    \left(\myPhi_{\myPsi} \myx\right)[n,j] = (\myx \ast \myw_j \ast \mypsi_j)[n].
\end{equation}
When initializing the filter entries of $\myPhi$ at random, e.g., $\myw_{j}\sim \mathcal{N}(0,\sigma^2\mathbb{1}_T)$, the hybrid encoder can be interpreted as a random filterbank with an inductive bias. This bias is inherited from the characteristics of $\myPsi$, and may embrace band limitation or a structured scale of center frequencies. By construction, these characteristics are also preserved during the optimization of $\myPhi_{\myPsi}$. If $\myPsi$ is an auditory filterbank, we call $\myPhi_{\myPsi}$ a \textit{hybrid auditory filterbank}.
Figure \ref{fig:filters} (right) illustrates the filters and their frequency responses of the hybrid auditory filterbank that we use for speech enhancement in Section \ref{sec:num}.

While the use of an auditory filterbank as encoder alone is already expected to be beneficial in speech--related tasks, the hybrid construction allows for data--driven fine-tuning, hence, further improvement of the alignment with the mask model.


\subsection{Stability of Hybrid Filterbanks and $\kappa$-penalization}\label{sec:stab}
A random filterbank with i.i.d. Gaussian weights forms a so-called \textit{random tight frame} \cite{ehler2015pre}, i.e., is tight in expectation \cite{haider2023randomfb},
\begin{equation}\label{eq:expPhi}
    \mathbb{E}\left[\Vert \myPhi\myx\Vert^2\right] = J T\sigma^2 \Vert\myx\Vert^2.
\end{equation}
A random hybrid filterbank $\myPhi_{\myPsi}$ inherits the stability properties of $\myPsi$ and $\myPhi$, and can be shown to also form a random tight frame.
\begin{thm}\label{thm:tight}
Let $\myPsi$ be a tight filterbank with frame bound $A_{\myPsi}$ and $\myPhi$ a random filterbank with length-$T$ filters. The associated hybrid filterbank $\myPhi_{\myPsi}$ is a random tight frame with
    \begin{align}
    \begin{split}
    \label{eq:random-tf}
    \mathbb{E}\left[\Vert \myPhi_{\myPsi} \myx\Vert^2\right] = A_{\myPsi}T\sigma^2 \Vert \myx\Vert^2.
    \end{split}
    \end{align}
\end{thm}
However, in any setting where the encoder filterbank is trainable, it is not guaranteed that it also remains stable during training. To counteract possible instabilities, for a given objective function $\mathcal{L}(\myx,\tilde{\myx})$ we propose to penalize the condition number $\kappa = B/A$ of $\myPhi$ by minimizing
\begin{equation}\label{eq:reg}
    \mathcal{L}_\beta(\myx,\tilde{\myx}) = \mathcal{L}(\myx,\tilde{\myx}) + \beta\cdot \kappa,
\end{equation}
with a scaling parameter $\beta>0$.
Note that minimizing $\kappa$ as above is less restrictive than minimizing $\Vert \mathbb{1} - \Phi^\top \Phi \Vert$ as proposed in \cite{cisse2017parseval}.
Furthermore, the computation of $\kappa$ can be done efficiently in the Fourier domain.
Denoting by $\widehat{\myw}_j$ the discrete Fourier transforms (DFT) of the filters $\myw_j$, which have been zero-padded to have length $N$, then $\myPhi^\top \myPhi$ is diagonalized as $\myPhi^\top \myPhi = U^\ast \Sigma U$,
where $\Sigma = \operatorname{diag}\left( \sum_{k=1}^J \vert \widehat{\myw}_k \vert^2 \right)$ and $U$ is the unitary DFT matrix.
Hence, the frame bounds coincide with the smallest and largest eigenvalue of $\Sigma$, given by
\begin{align}\label{eq:specbound1}
    A = \min_{0\leq k\leq N-1} \sum_{j=1}^J \vert \widehat{\myw}_j[k] \vert^2, \quad B = \max_{0\leq k\leq N-1} \sum_{j=1}^J \vert \widehat{\myw}_j[k] \vert^2.
\end{align}
From \eqref{eq:specbound1} we can deduce that the gradient of $\kappa$ is well-defined if the filterbank forms a frame.
Hence, using FFT methods it becomes feasible to include the computation of $\kappa$ and its gradient in iterative gradient-based optimization procedures for training.

In applications, convolution is usually performed with a stride to reduce redundancy, i.e., using a hop--size in the sliding filter (downsampling). This is inherent to the application of the filterbank and must be taken into account when calculating $\kappa$ in \eqref{eq:reg}. In the context of Eq. \eqref{eq:specbound1} this requires taking into account aliasing effects \cite{balazs2017framespsycho}. Assuming that these effects are negligible in our application for small stride values, we ignore aliasing in this work and leave a detailed discussion to future work.

\section{Model Implementation for Speech Enhancement and Training}\label{sec:num}
We demonstrate the proposed hybrid auditory filterbank and $\kappa$-penalization in a speech enhancement task, i.e., given a noisy signal $\myx_{\text{noisy}} = \myx + n$ we aim to suppress the noise signal $n$ via an encoder--mask--decoder model.

\begin{table*}[ht]
  \centering
  \begin{tabular}{@{}lllllll@{}}
    \toprule
    \rowcolor{lightgray}
    Encoder & Params & Objective & $\kappa$-penalization & PESQ & SI-SDR & \textbf{$\kappa$} \\
    \midrule
    \textbf{STFT} (baseline)& 0 & MCS & \ding{55} & 3.19 & 9.85 & 2  \\
    \textbf{audlet} (ours)& 0 & MCS & \ding{55} & 3.23 & 9.58 & \textbf{1} \\
    \midrule
    \textbf{conv1D}& 8.1k & MCS & \ding{55} & 2.66 & 11.69 & 3.2  \\
    \textbf{conv1D} & 8.1k & MCS$_\beta$ & \ding{51} & 2.77 & \textbf{11.99} & \textbf{1}   \\
    \textbf{hybrid audlet} (ours) & 2.8k & MCS & \ding{55} & 3.38 & 8.86 & 1.2 \\
    \textbf{hybrid audlet} (ours) & 2.8k & MCS$_\beta$ & \ding{51} & \textbf{3.39} & 8.68 & \textbf{1} \\
    \bottomrule
  \end{tabular}
  \caption{Speech enhancement benchmark on CHiME-2 WSJ-0. MCS: mixed compressed spectral loss \eqref{eq:mcs}. PESQ: perceptual evaluation of speech quality. SI-SDR: scale-invariant signal-to-distortion ratio. $\kappa$: condition number of the encoder (lower is better). }
  \label{tab:metrics}
\end{table*}

\subsection{Encoder/Decoder Design} \label{sec:enddec0}
We compare four different encoder configurations. Each one operates with 256 channels.
\begin{enumerate}
    \item \textbf{STFT} (baseline): A STFT with Hann window of length 512 and a hop-size of 256. The associated filterbank has a condition number $\kappa = 2$.
    \item \textbf{Audlet:} A tight auditory filterbank $\myPsi$ computed with the routine \texttt{audfilters} \cite{necciari2018audlets} from the LTFAT toolbox \cite{sondergaard2015ltfat} (Figure \ref{fig:filters} left). The filters are smoothed and cut to a length of 512 samples and a hop--size of 128 is used. This filterbank is comparable to a CQT with frequency-adaptive bandwidths and the center frequencies following the mel scale.
    \item \textbf{Conv1d:} A randomly initialized trainable filterbank $\myPhi$ with filters of length 32 and a hop-size of 8. This setting is reminiscent of the encoder used in Conv-TasNet \cite{lou2019convtasnet}.
    \item \textbf{Hybrid audlet:} A randomly initialized hybrid auditory filterbank $\myPhi_{\myPsi}$ composed of $\myPsi$ from 2. and a trainable filterbank $\myPhi$ from 3. with filters of length 11 and a hop-size of 1.
\end{enumerate}
For the baseline, the decoder is the inverse STFT. For all other cases, the decoder is the transposed filterbank of the encoder and is not optimized, i.e., the weights are shared. Using $\kappa$-penalization 
\eqref{eq:reg}, this will always be close to a dual for the encoder.
To benefit from fast convolution on GPU we implement all encoder decompositions via Pytorch's \texttt{conv1d} \cite{cheuk2020nnaudio}. Complex convolution is implemented separately regarding the real and imaginary parts. The results are shown in Figure \ref{fig:resp}.

\begin{figure}[t]
    \centering
    \includegraphics[width =\linewidth]{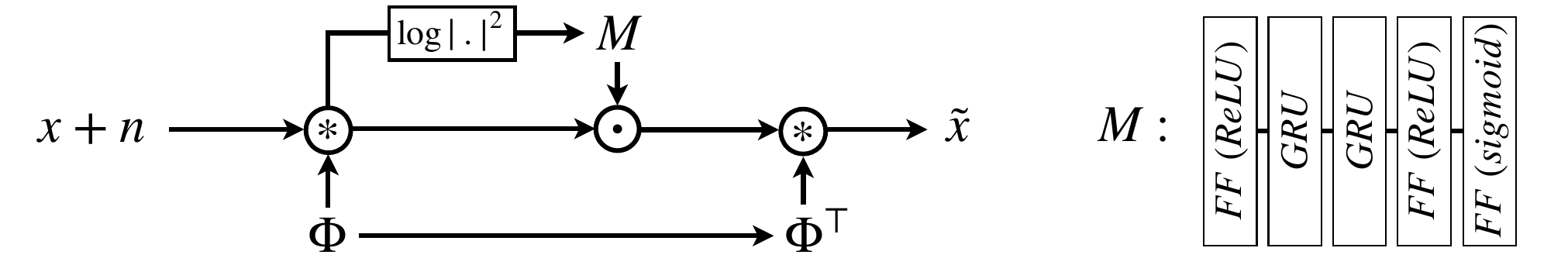}
    \caption{Left: Encoder--mask--decoder architecture: Encoder $\Phi$ (convolution), mask $M$ (point-wise multiplication), decoder $\Phi^\top$ (convolution and summation). Right: Mask model architecture consisting of feed-forward layers and gated recurrent units.}
    \label{Figure:Model}
\end{figure}

\subsection{Mask Model Architecture}
Based on the log magnitude responses of the encoder, the central part of the model computes a mask that is applied to the responses before being decoded. Following the simple and effective architecture proposed in \cite{xia2020} the mask consists of a feedforward layer, two GRU layers, and another three feedforward layers (Fig. \ref{Figure:Model}). The last layer uses a sigmoid activation function, all others are activated by a ReLU. In total, the mask model has 2.78m trainable parameters.

It should be noted that we apply the trainable filters before taking the log magnitude of the coefficients. This is to be distinguished from an architecture where a convolutional layer is applied on the log magnitude coefficients from a fixed filterbank.

\subsection{Training}
We adapt the mixed compressed spectral loss introduced in \cite{braun2020data} as our learning objective. Traditionally, this loss uses STFT coefficients, but we generalize it to the coefficients of our encoder filterbank $\myPhi$. Letting $\varphi$ and $\tilde{\varphi}$ denote the phases of $\myPhi \myx$ and $\myPhi \tilde{\myx}$ respectively, we perform empirical risk minimization with respect to
\begin{align}\label{eq:mcs}
    \text{MCS}(\myx,\tilde{\myx}) &= \gamma\cdot \big\Vert \vert \myPhi \myx\vert^c e^{j\varphi} - \vert \myPhi \tilde{\myx}\vert^c e^{j\Tilde{\varphi}} \big\Vert^2 \nonumber \\
    &+(1-\gamma)\cdot \big\Vert \vert \myPhi \myx \vert^c - \vert \myPhi \tilde{\myx} \vert^c \big\Vert^2.
\end{align}
For fixed encoders, we found that it is crucial to design the objective function based on the representation that is also used to estimate the mask. For trainable encoders, this representation, hence, the loss function changes with training. Following \cite{braun2020data}, we choose compression and weighting terms as $c=\gamma = 0.3$ which has been found to perform best for the proposed mask model in terms of the highest PESQ score~\cite{PESQ2001}.
When using $\kappa$-penalization described in \eqref{eq:reg}, we aim to minimize
\begin{align}
\text{MCS}_\beta(\myx,\Tilde{\myx}) = \text{MCS}(\myx,\Tilde{\myx}) + \beta\cdot \kappa.
\end{align}
By experimental exploration in our setting, we identified $\beta = 10^{-5}$ as a good value that is sufficiently small to not interfere with the minimization of the objective and sufficiently large to produce tightness consistently.

As optimizer, we use AdamW~\cite{loshchilov2018decoupled} with an initial learning rate of $10^{-4}$ and validate every 10 epochs. The batch size is 32. The model with the highest PESQ score on the validation set is selected for evaluation on an unseen test set. The performance of this model is reported in terms of PESQ and SI-SDR~\cite{roux2018sdr}.

\subsection{Dataset}
We use the CHiME-2 WSJ-0 dataset~\cite{Vincent2013} which consists of 7138 (train), 2418 (dev), and 1998 (test) speech utterances in English, from which we take \SI{5}{s} excerpts, respectively. The sampling rate is \SI{16}{kHz}. Every sample consists of a reverberate speech signal and a noise signal, added with an \gls{SNR} ranging from -6 up to \SI{9}{dB} in steps of \SI{3}{dB}. The target signal is the corresponding reverberated speech signal.



\section{Results and Discussion}

\subsection{General}
The main benefits of the proposed methods lie in the enhanced usability of trainable filterbanks in an encoder--decoder setting.
The fixed filterbank can be flexibly chosen to fit the problem at hand, and construction, implementation, and training using the proposed MSC loss is straightforward.
Enforcing tightness via $\kappa$-penalization provides the following:
\begin{itemize}
    \item the encoder output level is under control and easily adjustable
    \item the decoder does not have to be computed
    \item stability: small perturbations have small effects
\end{itemize}
In all our experiments, $\kappa$-penalization did not negatively influence the optimization of the main objective function, and we did not observe a noticeable loss in computational time. 

\subsection{Speech Enhancement}
The outcome of the speech enhancement task aligns very well with our expectations (c.f. Table \ref{tab:metrics}):
\begin{itemize}
    \item The audlet encoder outperforms the STFT in terms of PESQ.
    \item The hybrid audlet filterbank yields the highest PESQ overall, with a significant increase of $0.2$ compared to the baseline.
    \item Conv1d with random initialization yields the best SI-SDR.
\end{itemize}
Not only does the hybrid audlet filterbank outperform all other models in some aspects, it also reaches an optimal condition number at the end of training due to $\kappa$-penalization. We note that on the relatively short trainable filters it has only limited effect. For conv1d, the effect is larger ($\kappa=3.2$). Although not significantly, $\kappa$-penalization yields better scores in all the cases. 

Conv1d takes a long time to train and is very sensitive to hyperparameters such as filter length, stride, learning rate, and $\beta$. On the contrary, the hybrid filterbanks learn fast and work in many hyperparameter configurations.
We conjecture that the high SI-SDR score by conv1d comes from the fact that gradient descent treats every filter equally, such that the contributions of the filters average out over the different bands. As a result, the MCS resembles an energy measure, visible in the center plot of Figure \ref{fig:resp}.  

\subsection{Limitations and Outlook}
Using hybrid filterbanks does not speed up inference compared to the baseline. However, the related work on multiresolution neural networks paves the way towards reducing the number of parameters and saving computations.

With regard to the evaluation metrics provided in Table \ref{tab:metrics}, it can be observed that the effect of $\kappa$-penalization is relatively minor. It remains to be seen whether the benefits of stabilization are comparably more significant in cases where the condition number tends to grow in the absence of explicit penalization in the learning objective. 

\section{Conclusion}
This paper presents three methodological contributions to (hybrid) filterbank learning for speech enhancement. Firstly, we design trainable hybrid encoders for audio feature extraction with desirable properties, such as band limitation, and fixed center frequencies of the filters. The properties can be set in advance and persist throughout training. Secondly, a frame theoretic perspective provides the theoretical backbone for defining a simple and effective stabilization mechanism that keeps any trainable filterbank very close to tight throughout training. The implications are that the filterbank is norm preserving and can be inverted by its transpose. The third point is the adaption of the mixed compressed spectral loss to the encoder coefficient domain. In a speech enhancement task, this framework manages to outperform the performance of the STFT and randomly initialized conv1d layers in terms of the PESQ score.
While this contribution focuses on demonstrating the methods in the specific application of speech enhancement, in future work we will advance the theory and extend the experiments to other tasks and domains to outline the universality of the approach.

\section{Acknowledgments}
D. Haider is recipient of a DOC Fellowship of the Austrian Academy of Sciences at the Acoustics Research Institute (A 26355). V. Lostanlen is supported by ANR project MuReNN (ANR-23-CE23-0007-01). The work of P. Balazs was supported by the FWF projects LoFT (P 34624) and NoMASP (P 34922). 
The authors would particularly like to thank Clara Hollomey for making audlet filterbanks available in Python.

\bibliographystyle{IEEEtran}
\bibliography{mybib}

\end{document}